# Holographic Cell Stiffness Mapping Using Acoustic Stimulation

Rahmetullah Varol[1], Sevde Ömeroğlu[2], Zeynep Karavelioğlu[3], Gizem Aydemir[4], Aslıhan Karadağ[5], Hanife Ecenur Meço[6], Gizem Çalıbaşı Koçal[7], Muhammed Enes Oruç[8], Gökhan Bora Esmer[9], Yasemin Başbınar[10], Hüseyin Üvet[*,11]

[1]Yildiz Technical University, Mechatronics Engineering Department, 34349 Istanbul/Turkey

[2]Gebze Technical University, Chemical Engineering Department, 41400 Kocaeli/Turkey

[3]Yildiz Technical University, Bioengineering Department, 34220 Istanbul/Turkey

[4]Sorbonne Université, Institut des Systèmes Intelligents et de Robotique, 75005 Paris/France

[5]Dokuz Eylul University, Oncology Institute, 35330 İzmir/Turkey

[6]Dokuz Eylul University, Oncology Institute, 35330 İzmir/Turkey

[7]Dokuz Eylul University, Oncology Institute, 35330 İzmir/Turkey

[8]Gebze Technical University, Chemical Engineering Department, 41400 Kocaeli/Turkey

[9]Marmara University, Electrical and Electronics Engineering Department

[10]Dokuz Eylul University, Oncology Institute, 35330 İzmir/Turkey

[11]Yildiz Technical University, Mechatronics Engineering Department, 34349 Istanbul/Turkey

*Corresponding author: Huseyin Uvet (huvet@yildiz.edu.tr)

# Abstract

Accurate assessment of stiffness distribution is essential due to the critical role of single cell mechanobiology in the regulation of many vital cellular processes such as proliferation, adhesion, migration, and motility. Cell stiffness is one of the fundamental mechanical properties of the cell and is greatly affected by the intracellular tensional forces, cytoskeletal prestress, and cytoskeleton structure. Furthermore, it has been shown that alterations in mechanical properties are associated with the pathogenesis and progression of various diseases. Measurement of cell stiffness plays a crucial role in understanding the metastasis and differentiation mechanisms of cancer. Herein, we propose a novel holographic single-cell stiffness measurement technique that can obtain the stiffness distribution over a cell membrane at high resolution and in real-time. Periodic acoustic pressure waves stimulate the cell membrane, and the membrane deformation is measured via a digital holographic microscope. The proposed imaging method coupled with acoustic signals allows us to assess the cell stiffness distribution with a low error margin and label-free manner. We demonstrate the proposed technique on HCT116 (Human Colorectal Carcinoma) cells and CTC-mimicked HCT116 cells by induction with transforming growth factor-beta (TGF-β). Validation studies of the proposed approach were carried out on certified polystyrene microbeads with known stiffness levels. Its performance was evaluated in comparison with the AFM results obtained for the relevant cells. When the experimental results were examined, the proposed methodology shows utmost performance over average cell stiffness values for HCT116, and CTC-mimicked HCT116 cells were found as 1.08 kPa, and 0.88 kPa, respectively. The results confirm that CTC-mimicked HCT116 cells lose their adhesion ability to enter the vascular circulation and metastasize. They also exhibit a softer stiffness profile compared to adherent forms of the cancer cells. Hence, the proposed technique is a significant, reliable, and faster alternative for in-vitro cell stiffness characterization tools. It can be utilized for various applications where single-cell analysis is required, such as disease modeling, drug testing, diagnostics, and many more.

# Introduction

Cell mechanics is highly dynamic and regulated by structural, cross-linking, and signaling molecules (Fletcher & Mullins, 2010). Studies of various diseases using different experimental techniques have shown that changes in cancer cells' mechanical properties are associated with disease pathogenesis and progression (W. Xu et al., 2012). The mechanical properties of cells are mainly defined by the actin cytoskeletal structure (Efremov et al., 2014). This cytoskeleton develops during the differentiation of cells, plays a role in many cellular functions, and undergoes various characteristic transformations during many diseases, including cancer (Li et al., 2008; Vasiliev, 2004).

The mechanical behavior of cells is closely related to the cellular architecture and regulation of biological functions such as proliferation, migration (Bangasser et al., 2017), motility and cell adhesion properties. Therefore, changes on the mechanical properties of cells indicate the alterations on the biological phenotype. Studies conducted on different diseases by using a variety of experimental techniques have shown the alterations on mechanical properties associated with pathogenesis and progression of the diseases (Wala & Das, 2020). Cell stiffness is a mechanical parameter that is closely related to intracellular tensional forces, cytoskeletal prestress and cytoskeleton structure and affects the fundamental activities of the cells. Accurate measurement of a cell's stiffness enables the assessment of various properties of a cell such as viability, proliferation, differentiation, migration, and invasion which are involved in the disease development.

The transformation of cancer cells from normal to malignant form is also characterized by changes in the mechanical properties (Kumar & Weaver, 2009). It has been shown in various studies that cancer cells tend to have a lower mechanical stiffness value compared to normal cells (Suresh, 2007). It has also been demonstrated that metastatic cells exhibit a softer profile than benign cells (Cross et al., 2007). Moreover, it is accepted that cancer cells send signals in the form of secreted extracellular vesicles that can affect the mechanical conditions of other cells (Zanotelli et al., 2017). Hence, it is essential to examine the cell's mechanical properties at the single cell level in more detail to understand the cancer progression mechanism better.

Measurement of cell stiffness is an extensively studied topic, and many advanced techniques have been utilized for this purpose. These techniques can be categorized broadly into three different classes: subcellular strategies, whole-cell measurement techniques, and multicellular measurement techniques. Generally, multicellular methods allow the faster analysis of cell populations while sacrificing the captured signals' resolution. On the other hand, subcellular strategies enable the analysis of subcellular components (such as membrane or organelles), but they usually require a significant analysis duration.

The most widespread method is the use of atomic force microscopy (AFM), which can be categorized under both the subcellular measurement techniques and whole-cell measurement techniques depending on the number of points on which the cell membrane was indented. This technique is based on applying a predetermined force on the cell membrane via the AFM cantilever. Deformation on the membrane is measured by the deflection of the cantilever using a laser and a photodetector. Captured information is subsequently used to determine the stiffness of the cellular membrane. AFM based technique was previously used for the determination of different mechanical characteristics of live and dead cells (Lulevich et al., 2006), stiffness levels of breast cancer cells (Li et al., 2008), and morphological and mechanical changes induced by the application of various substances (Jin et al., 2010). Interpretation of AFM results is usually made according to contact mechanics in which different deformation models are used to relate the applied forces with the expected deformation. One of the most commonly used models is the Hertzian model, which gives good results for small deformations (Kontomaris, 2018).

Another commonly applied technique is the use of optical trapping for whole-cell stiffness measurement. Optical trapping allows the measurement of the Young's modulus of single cells through a dual-beam optical trap to induce mechanical stress on whole cells. One advantage of this method is that it can be used on suspended cells (Guck et al., 2005).

There are several more sophisticated techniques, including magnetic twisting cytometry (Aermes et al., 2020), deformation cytometry, particle-tracking microrheology (Wirtz, 2009), micropipette aspiration (Trickey et al., 2006), parallel-plate rheometry, cell monolayer rheometry, optical stretching (Dao et al., 2003), and micro-robotic probing. However, there are still more grounds to cover as these techniques usually yield varying results up to 1000-folds (Wu et al., 2018). Furthermore, many of these methods typically rely on the samples' heavy preconditioning, such as fixation or attachment of various surface markers. These procedures compromise the viability of cells and do not allow for repeatable measurements over a single sample.

Here, we present an acoustic simulation-based technique where surface acoustic waves generated by a lead zirconate titanate (PZT) acoustic transducer are applied through a cell medium. The cells' mechanical response is measured through the use of digital holographic microscopy (DHM). A constant frequency acoustic signal is applied, and the dynamic response of cells is measured in high frequency by sampling small patches of the interferogram captured by a complementary metal-oxide-semiconductor (CMOS) sensor. The proposed technique paves the way to obtain shorter analysis time and higher resolution with a low error margin compared to equivalent methods. Furthermore, the proposed approach can be used to analyze the cells without directly contacting the cells. Thus, cell viability is not damaged by the possible cell-probe interactions, and cells can be examined in their environment. Reading cell biomechanics within the aforementioned advantages will support this study to be actively involved in diverse biological studies such as cancer research and drug efficacy evaluation.

**Significance:** This study's main contribution is the development of a novel label-free and contact-free methodology for the measurement of the stiffness distribution over a cellular membrane using acoustic stimulation and holographic imaging. It is much faster and practical than traditional cell stiffness measurement or mapping methods, particularly compared to the AFM-based measurement technique, which is commonly considered the current gold standard methodology. Consequently, higher resolutions of stiffness distribution can also be achieved under higher magnification ratios. More importantly, the proposed method presents precise assessments and has a low error margin. Besides, the possible mechanical interactions with the single-cell can be eliminated, and in this way, the cells' viability can be preserved. The ability to scrutinize the cell stiffness with all these advantages and analyze it comparatively with other studies conducted on the cytoskeleton and apoptosis can shed light on cancer studies, neurodegenerative disease research, and drug efficacy tests, so this technique will get involved in several biomedical applications.

# Results

**Experimental setup**

The experimental setup in this study consists of a PZT transducer that was used to generate bulk acoustic waves inside the fluidic chamber. Bulk acoustic waves cause an acoustic pressure on the surface of the cells which in turn deforms the cell membrane proportional to the acoustic force. We observe this displacement pattern using a high-speed camera and record the dynamic response of the cell membrane to the acoustic source over a frequency range. The acoustic signal that is used to drive the PZT transducer (Steminc SMBA25W73T05PV) is generated by an arbitrary waveform generator (Siglent SDG6032X). The signal is then amplified by a voltage amplifier (Falco Systems, WMA-300) and fed into the transducer. Recording is done using a Mach-Zehnder interferometric setup, so that the morphological transformation of the cell can be determined with high accuracy. Our holographic imaging setup is based on a phase shifting inline Mach–Zehnder interferometer. A schematic view of the setup is given in Figure 1-(a). As a coherent light source, we used a 527 nm, 10 mW He-Ne laser. A high-frequency piezo actuator (New Focus, Irvine, CA, USA, Picomotor 8302) actuated in 1000 steps per second was used for phase-stepping. We developed an algorithm for real-time holographic reconstruction using a continuously shifting piezo actuator and a high-speed CMOS camera (ZEISS, Oberkochen, Germany, AxioCam 702 mono). The CMOS camera was synchronized to the motion of the piezo actuator and at each step captured an image. A 20x objective (Newport, M-20X) was used during the experiments. The acoustic setup that is attached to the fluidic chamber is shown schematically in Figure 1-(b) along with the computer-aided design view of the experimental setup in Figure 1-(c).

**Acoustic stimulation and holographic reconstruction**

A 1 kHz thickness mode acoustic transducer is used to generate the acoustic waves used to drive the vibration motion of the cells. The transducer's vibration motion creates propagating surface waves over the glass slide, which in turn generates a vibration motion on the cells. We applied vibration frequencies in the range from 50 Hz to 1000 Hz with 50 Hz increments for each sample. We assume that the cell is oscillated by a periodic stimulus of frequency $f_0$. In order to capture the interferograms at desired points throughout the vibration period, we trigger the camera using a periodic signal of frequency $f_0/(1 + f_0/\delta_t)$. After $(\delta_t/f_0)$ cycles, we obtain a set of interferograms for $k\delta_t$ time-shifts for $k = 0, 1, \ldots, n$ where $n\delta_t = 1/f_0$. This process was repeated for each phase-shift, which gives us a set of phase-shifted interferograms for each time step.

**Temporal alignment of captured interferograms**

To match the set of interferograms for each phase shift, we use an algorithm that gives a measure of how likely a given interferogram is the phase-shifted pair of another interferogram. Using this algorithm, we determine how many frames should be shifted in the second set of interferograms to obtain the matched frames of the first and second set. Repeating this process for each phase-shift gives us the correct matching set of interferograms. Reconstructing each of these sets provides us a set of surface morphologies sampled periodically throughout a period of the vibration movement. For each cell, we determine edges in a given depth map using a watershed marker-controlled segmentation technique. Using the set of depth data, we calculate the mean displacement value for each pixel within the cell's boundaries.

**Microbead stiffness measurement**

The first set of experiments were conducted on reference microbeads (Thermo Scientific Duke Standards, 2005ATS). These are polystyrene microbeads that have an average diameter of 5 $\mu$m and an average stiffness of 1.05 GPa. They were used to calibrate and validate the stiffness measurement obtained from the acousto-holographic imaging system. The processed holographic image of the microbeads can be seen in Figure 2-(a). The corresponding depth map of the microbead is shown in Figure 2-(b). Depth measurements were performed at 500 Hz, and the periodic vibration of the beads was constructed from these measurements. An illustration of the collected peak depth measurements and the sinusoidal vibrational signal obtained from these measurements can be seen in Figure 2-(c). Subsequently, amplitude and frequency parameters were obtained from that sinusoidal signal to determine Young's modulus parameter for each microbead. The stiffness distribution obtained from a single microbead compared to the analytically calculated distribution is shown in Figure 2-(d). Micorbead measurement results are summarized in Figure 3. To test the reliability of the system, reference microbeads, whose stiffness levels are known in the literature, have been used. When we compared the obtained results with the acousto-holography method, the stiffness degrees of microbeads were found as 1.05 GPa, and these values are in accordance with the values given in the literature.

**Cell stiffness measurement**

A similar procedure was then applied for the HCT116 and CTC-mimicked HCT116 cells. Stiffness distributions of the cells are given in Figure 4. Average cell stiffness values for HCT116 and CTC-mimicked HCT116 cells were found to be 1.08 kPa, and 0.88 kPa, respectively. Furthermore, AFM measurements on epithelial HCT116 cells were conducted, and the resulting average cell stiffness value (1.16 kPa) was compatible with the result obtained from the acousto-holographic system.

**TGF-$\beta$ treatment of HCT116 cells**

Transforming growth factor-beta (TGF-$\beta$), is a multifunctional cytokine that is commonly used for inducing the epithelial-mesenchymal transformation (EMT) in various epithelial cell lines (J. Xu et al., 2009). During this process, the epithelial characteristics deteriorate, and the cells acquire a migratory behavior. Deterioration of the epithelial characteristics includes dissolution of cell-cell junctions, adherent junctions, desmosomes, and loss of epithelial cell polarity. After which cells acquire a mesenchymal phenotype, characterized by actin reorganization and stress fiber formation, migration, and invasion. TGF-$\beta$ signaling in the body is one of the main precursors of metastasis and is shown to increase the formation of circulating tumor cells (CTCs) (Xie et al., 2018). In order to demonstrate the ability of the proposed acousto-holographic methodology to analyze the effects of EMT on epithelial cells, we treated the HCT116 cells with TGF-$\beta$ and obtained CTC-mimicked HCT116 cells. Epithelial cells were seeded in the fluidic chamber and incubated for 24 hours. Subsequently, cells were treated with TGF-$\beta$ and observed after a 48-hour incubation period. The results are shown in Figure 4-(c).

When the results obtained were examined, it was seen that HCT116 cells had a stiffer profile while HCT116 cells treated with TGF-$\beta$ were shown to have a softer structure compared to non-stimulated HCT116 cells. As the cells transition to a malignant state, their cytoskeletal structures change from an organized network to an irregular network, and this transformation then reduces the stiffness of the cells. They usually have a smaller number of tensile fibers; residual microfilament bundles are irregular, and maturation of focal adhesions is impaired. As a result, it has been found that tumor cells are softer than benign and healthy ones. This may be associated with changes in cell mechanics of the cancer cells to metastasize. Results of validation studies for confirmation of mesenchymal character induced by TGF-β treatment are shown in Figure 5.

**AFM indentation experiments**

AFM indentation technique is one of the most commonly used techniques for determining the stiffness distribution over a cell surface. In order to compare the results obtained with the acousto-holographic method, we performed a series of AFM indentation tests on epithelial HCT116 cells. These measurements were conducted using a stand-alone commercial AFM (Bruker Dimension Icon

AFM). A triangular silicon nitride cantilever (Bruker, DNP-10), with a nominal force constant of 0.06 N/m and a nominal tip radius curvature of 20 nm, was used. The AFM probe was calibrated prior to the experiments using the device software and the spring constant was determined from the thermal noise spectrum (Hutter & Bechhoefer, 1993). The experiments were conducted on the force modulation mode at a frequency of 10 Hz. The cantilever tip was initially placed just next to the cell for each sample, and then force-displacement curves were obtained for 0.5 $\mu m$ intervals at a 25×25 $\mu m$ region. In total, 2601 points on the cell surface were indented. In the form of Young's moduli, results obtained from the AFM experiments are shown in Figure 3-(b). Young's moduli values are calculated by applying the Hertz model on the linear region of the force-indentation curve. AFM measurement of CTCs requires more complex localization and isolation techniques that can alter the mechanics of the cell due to their weaker adhesive properties (Deliorman et al., 2020). Therefore, AFM experiments were not repeated on the CTC mimicked HCT116 cells.

Our results indicate that different stiffness values of the cells can be explained by the differences in the cytoskeletal structures that give these cells their characteristic features, and the situation of the cells displaying a stiffer profile can be explained by the fact that these cells display a more granular structure (Nikolaev et al., 2014). Furthermore, the volume of the cell may decrease dramatically due to the progression of cellular death. Thus, the stiffness change and cell volume difference occurring between the same cell lines can be explained by the different life cycles of these cells. Focal adhesions; act as an interface between the cytoskeleton and the cell environment and transmit the mechanical force across the cell membrane. The force exerted by the cytoskeleton deforms the extracellular-matrix (ECM) structure, and thus a response occurs through the mechanical stiffness of the cell (Fusco et al., 2017). When the obtained results we obtained are examined, it is seen that the change in stiffness between cells can be explained by the shift in ECM structure, which is the result of changes in the form of the cytoskeleton caused by cellular stress and the change in cellular stiffness accordingly.

## Discussion

Observation of acoustic bulk wave compression of cells using a holographic imaging setup allowed us to obtain the cell stiffness distribution at a very high-resolution. The measurement throughput is limited by the field of view of the assembled optical configuration. The resolution in the stiffness distribution is determined by the used magnification ratio and the employed holographic imaging apparatus's optical limitation. Higher-resolution rates may allow measurement of the stiffness of subcellular structures. Most of the currently used techniques allow cell stiffness determination only at a single cell level and fail to capture the subcellular changes. Generally, methods that can be used for subcellular cell stiffness measurement have a low resolution, usually in the order of a few micrometers (Lam et al., 2012). To the best of the authors' knowledge, this is the first methodology that allows the untethered measurement of cell stiffness at a sub-micron resolution. The cells' viability is not compromised since the measurement is done in an untethered manner using only low power acoustic signals. Furthermore, integrating the proposed methodology into an incubator compatible system would allow the observation of cellular stiffness changes for extended intervals.

The feasibility and accuracy of the proposed methodology are demonstrated using reference microbeads with a predetermined stiffness distribution. Obtained results were highly correlated with the known values of bead stiffness (R=0.976). Furthermore, we present the proposed technique's ability to measure the stiffness distribution of live cells by comparing the stiffness results obtained by using the AFM indentation technique and proposed technique on HCT116 cells. The results show that the proposed approach's resolution was orders of magnitude higher than that of AFM results. Also, the preparation and measurement duration in the AFM experiments were much longer than the proposed method. Furthermore, the AFM indentation technique requires mechanical contact with cells, which reduces their viability.

To demonstrate an application in which the proposed technique would be beneficial, we treat epithelial HCT116 cells with TGF-$\beta$ and induce the EMT process, which increases their metastatic character. Analyzed results show that the stiffness of endothelial HCT116 cells is higher than the CTC-mimicked HCT116 cells. This result is expected due to the stiffer and more regular distribution of actin filaments in HCT116 cells than CTC-mimicked HCT116 cells.

Here, we present the differences in cell stiffness between those in a different metastatic phase using a newly developed acousto-holographic methodology to detect the displacement and vibration values that occur when cells are stimulated by an acoustic signal with high resolution and low margin of error. When the cell stiffness distributions (Figure 3) were analyzed, it can be seen that the cell stiffness of endothelial HCT116 cells is higher than the CTC-mimicked HCT116 cells that lose their adherent properties to metastasize. Cell stiffness is expected to be higher due to the stiffer and more regular distribution of actin filaments in HCT116 cells compared to CTC-mimicked HCT116 cells. This technique, which can perform faster and higher resolution analysis than AFM, will become a valuable tool for cancer research and drug efficacy tests based on the single-cell analysis.

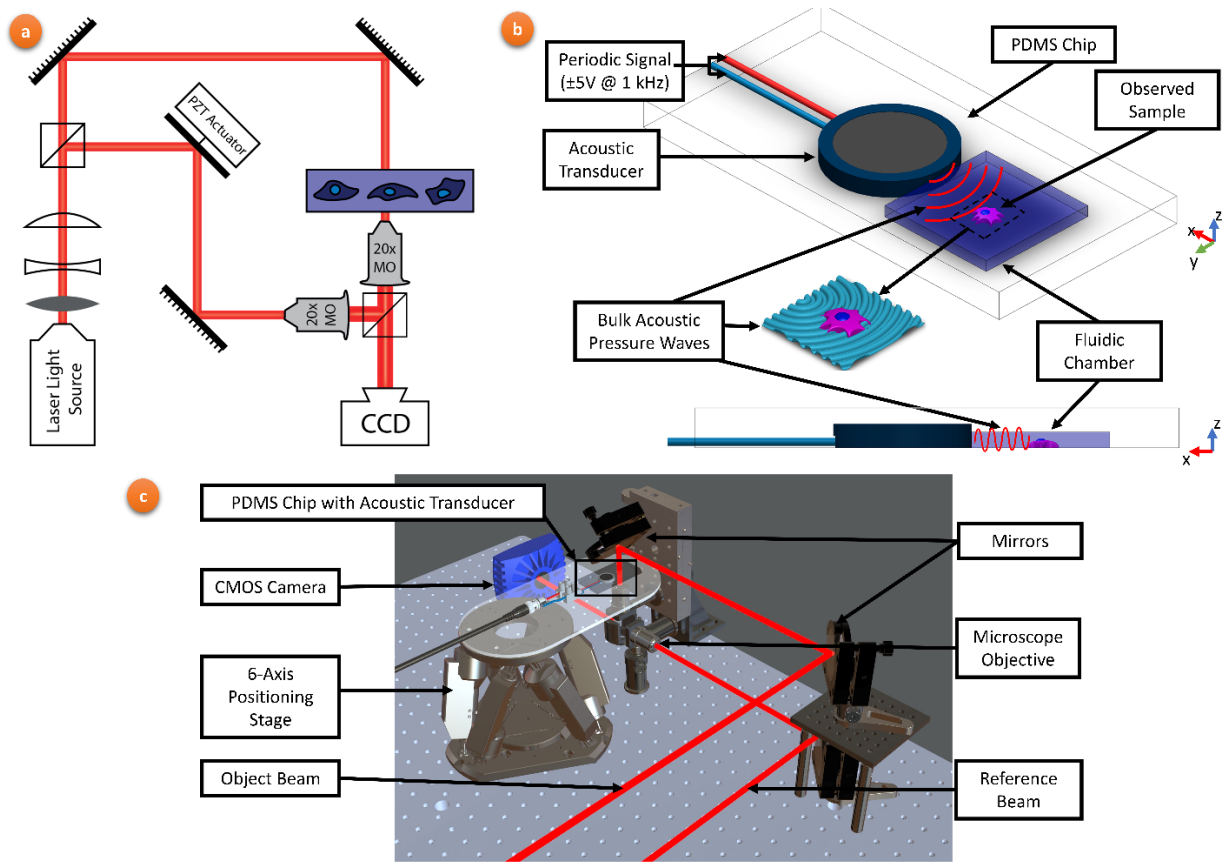

**Figure 1: a** Schematic representation of the holographic imaging setup which was based on a Mach-Zehnder interferometric setup. The setup contained a 527 nm He-Ne laser light source, a polarizer, a high-frequency piezo actuator actuated in 1000 steps per second that is used for phase-stepping, a high-speed CMOS camera and a 20x objective. **b** Shows the fluidic chamber that was used for creating the high-frequency acoustic signals which were used to stimualte the cells. The fluidic chamber consisted of a PDMS layer bonded on a glass layer and the bulk acoustic waves were generated by a PZT tranducer that was placed between the PDMS layer and the glass surface. **c** Shows a computer-aided design view of the experimental setup. A 6-axis position comtroller was used to manipulate the sample and the object beam propated through the fluidic chamber in which the stimulated cells resided. Then the object beam and reference beam was combined and directed into the CMOS sensor.

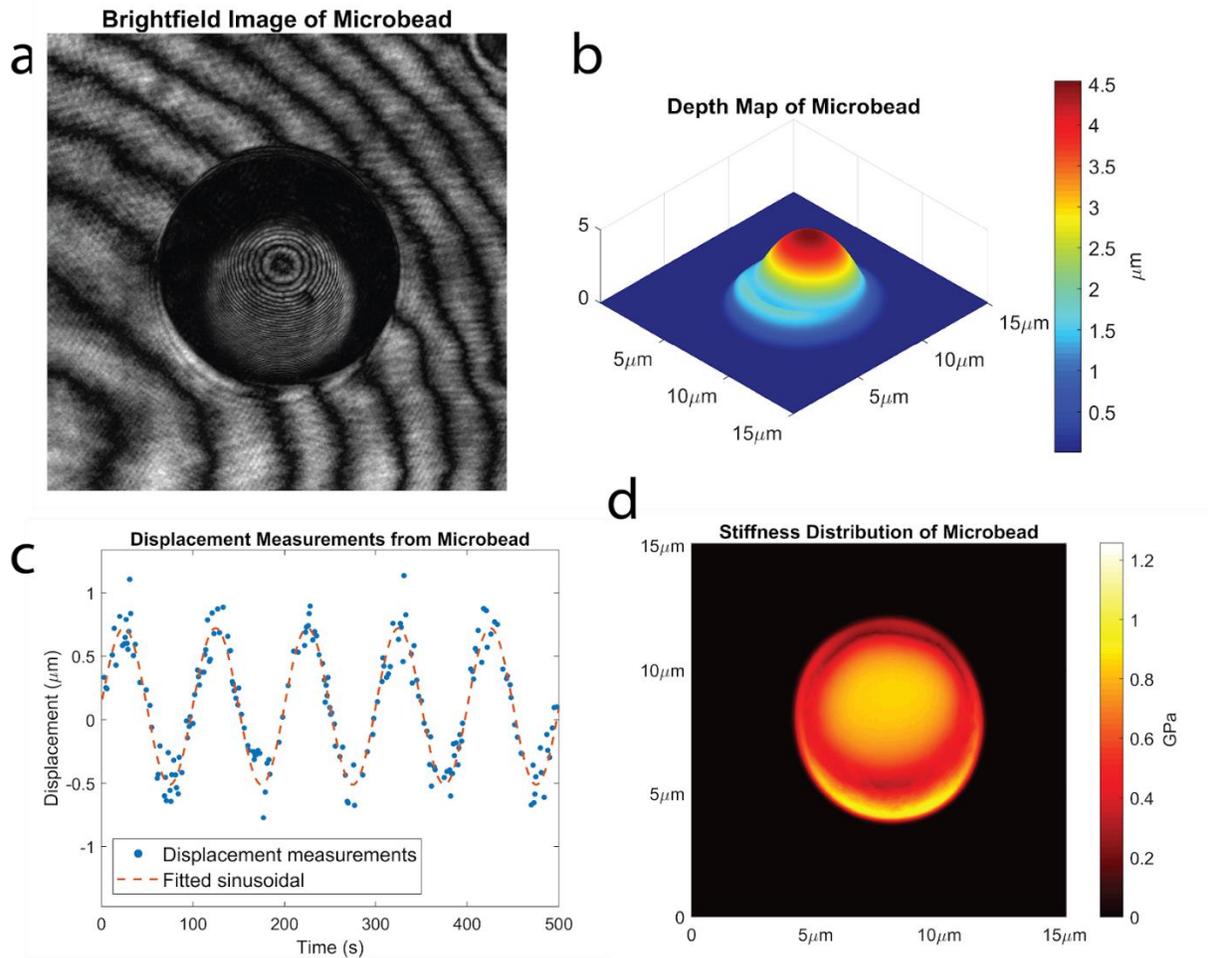

**Figure 2:** Stiffness measurement of microbeads that were used for calibration. **a** shows the amplitude image of the microbead. This image is obtained by resolving the real part of the diffraction equation. **b** shows the depth map of the microbead. This information is obtained through the reconstruction of the interferograms that were captured using the holographic imaging setup. By measuring the depth of a certain point on the surface of the object, the displacement graph **c** can be obtained. These displacement measurements can be used to determine the model parameters which yields the Young's modulus for each point on the surface of the object as shown in **d**.

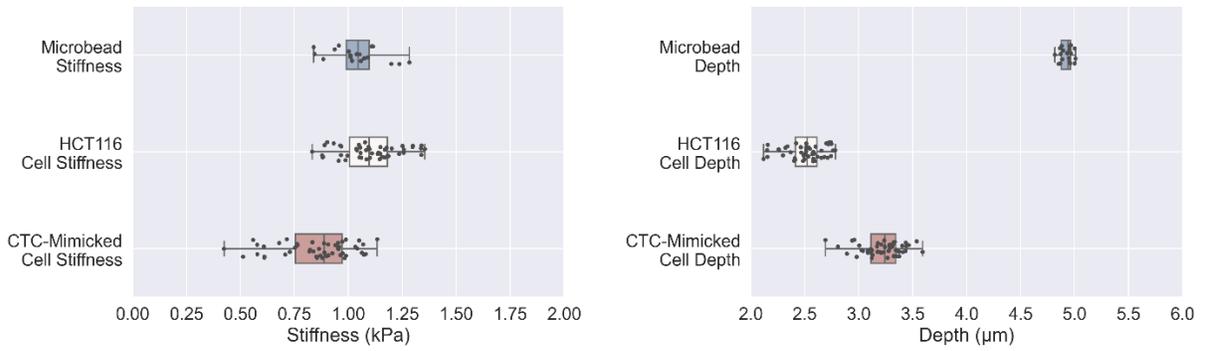

**Figure 3:** Average stiffness and depth measurements for microbeads, HCT116 cells and CTC mimicked HCT116 cells. Microbead stiffness was repoted in GPa while cell stiffness values were reported in kPa and all depth measurements were reported in $\mu m$s. It was observed the obtained microbead stiffness and depth measurement were on par with the reference values specified in the product datasheet which were 5 $\mu m$ for the microbead depth and 1.05 Gpa for the microbead stiffness. Furthermore, it was observed that CTC-mimicked cells had a lower stiffness value compared to their epithelial counterparts; one-way ANOVA comparison yielded p<0.05.

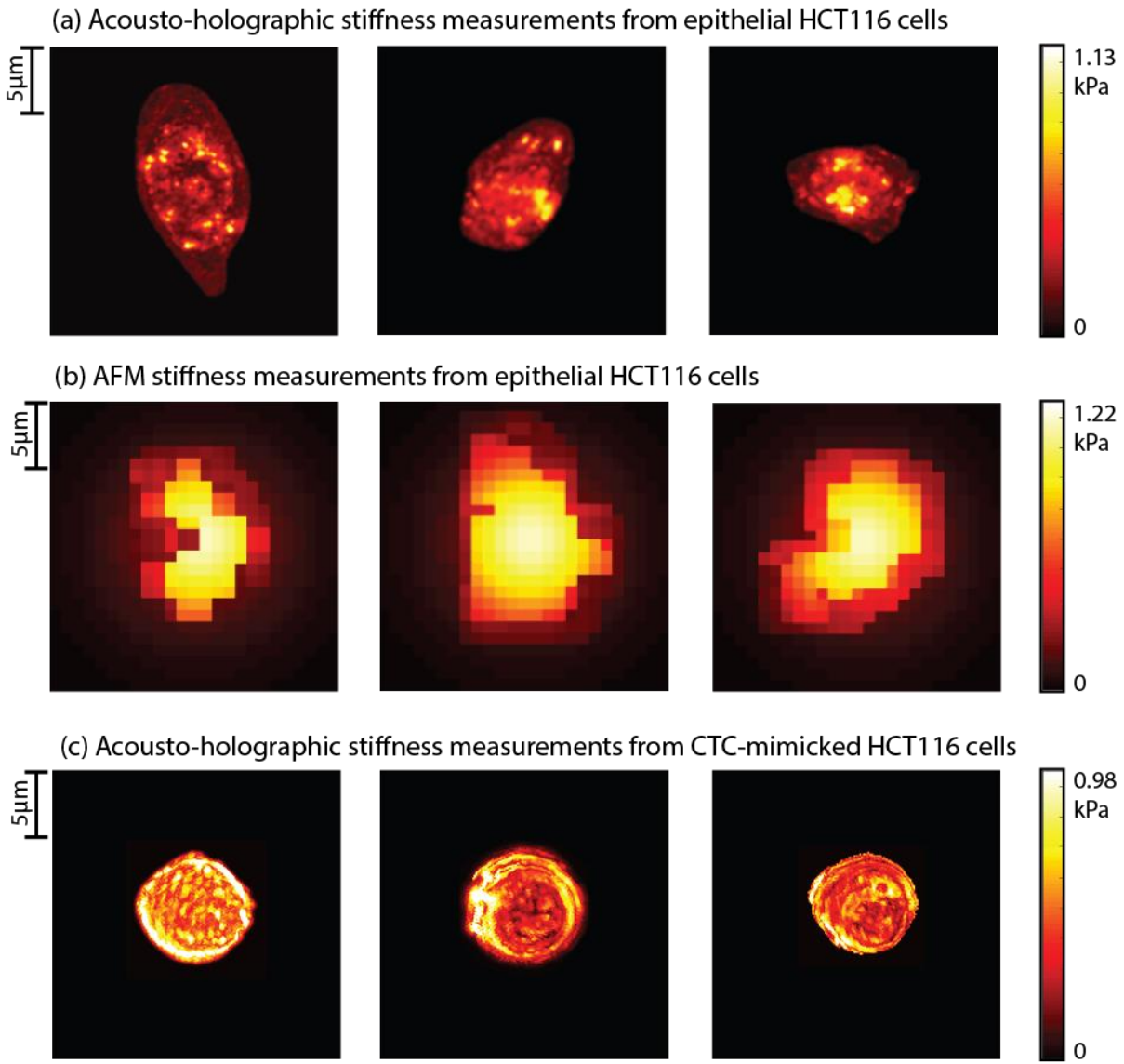

**Figure 3:** Stiffness distributions of the observed cell lines. It can be seen that the cell stiffness of endothelial HCT116 cells is higher than the CTC-mimicked HCT116 cells that lose their adherent properties to metastasize. Cell stiffness is expected to be higher due to the stiffer and more regular distribution of actin filaments in HCT116 cells compared to CTC-mimicked HCT116 cells.

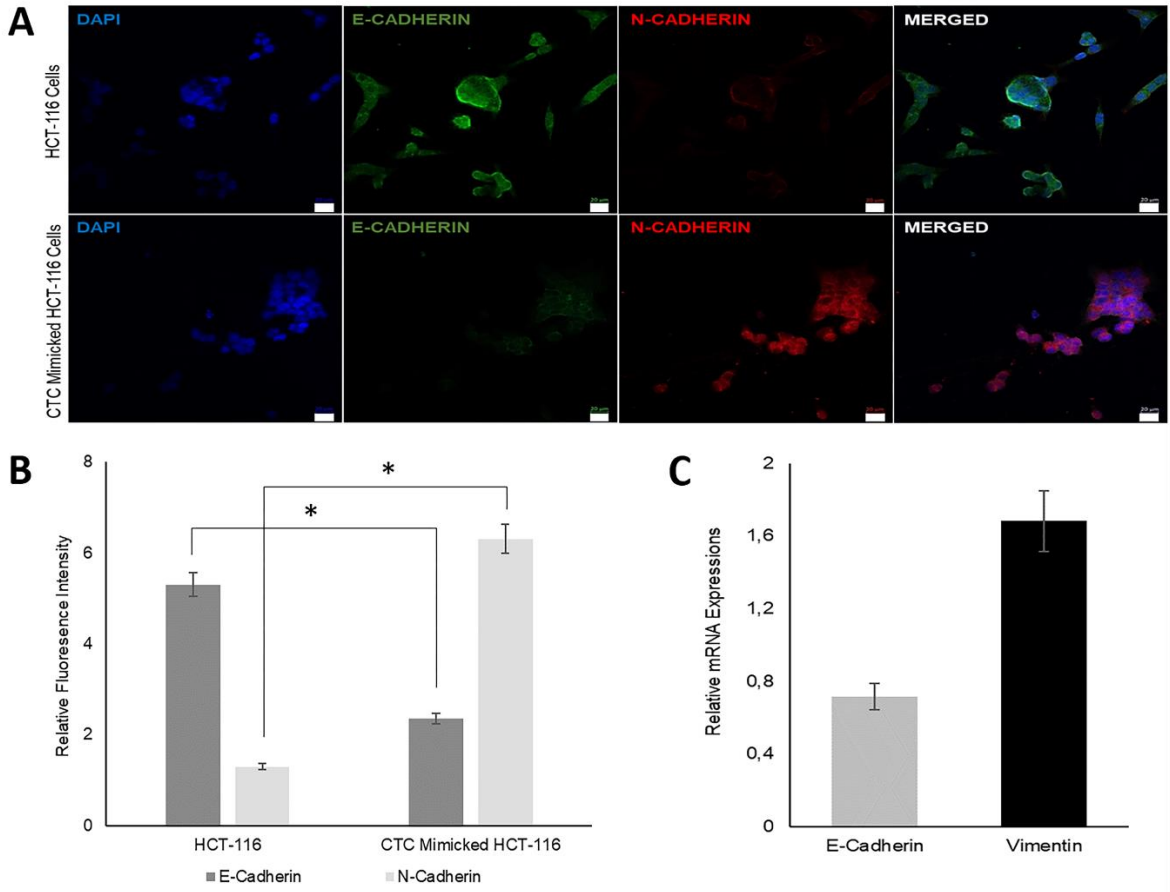

**Figure 5:** Validation of mesenchymal character induced by TGF-β treatment. A) Fluorescence immunocytochemistry staining for Dapi (blue), E-cadherin (green) and N-cadherin (red) on HCT116 and CTC mimicked HCT116 cells. Scale bar 20 μm. B) Bar graphs demonstrating the intensity of expressed E-cadherin and N-cadherin. Quantification of fluorescence intensities from arbitrary images of each condition was done with ImageJ software (NIH, Bethesda, MD, USA). All data were expressed as mean ± S.E.M., n = 3, star (*) indicates, $p < 0.05$. C) E-cadherin and Vimentin mRNA expression relative to housekeeping gene GAPDH by 2^-ddCt value calculation. E-cadherin expression was decreased 0.7-times and Vimentin expression was increased 1.7-times after TGF-β treatment for 48 hours, $p < 0.05$.

# Methods

## Fluidic Chamber

The fluidic chamber was fabricated on a microscope cover glass slide (0.13-0.17 mm thickness) using Sylgard 184 elastomer kit (PDMS). The PDMS base and curing agent were mixed in a ratio of 10:1 (w/w) and kept in the vacuum to remove air bubbles for 30 min. After degassing, PZT transducer was placed on the polymer and PDMS is cured in the oven at 65 °C for 3 hr. A fluidic chamber (12x12 mm) was created at the end of the piezoelectric transducer (PZT) placed in the PDMS structure. Oxygen plasma was applied at 400 mTorr pressure and high power for 2 minutes to irreversibly connect the created PDMS structure to the microscope glass slide. For the sterilization of the fluidic chamber, it was treated with ethanol in a sonicator for 5 min.

## Cell Culture

The human colorectal adenocarcinoma (HCT116, ATCC® CCL-247) was obtained from the American Type Culture Collection- ATCC (Manassas, VA, USA). These cells were propagated and maintained in McCoy's 5A (Cegrogen, Germany) cell culture media supplemented with 10% fetal bovine serum (Cegrogen, Germany), 1% penicillin/streptomycin (Cegrogen, Germany) and 1% L-glutamine (Cegrogen, Germany) in humidified incubator at 37°C with 5% $CO_2$. Then, HCT116 cells were treated with the transforming growth factor beta (TGF-β) (10 ng/mL) to stimulate CTC-mimicked cells displaying a mesenchymal phenotype.

## Holographic numerical reconstruction

The intensity distribution of a given interferogram obtained from the Mach-Zehnder interferometer can be expressed as follows:

$$I_{ij} = A_{ij} + B_{ij} \cos(\theta_j + \delta_j)$$

represents the $i$th phase-shifted interferogram ($i = 1,2,…,M$) and $j$ denotes the individual pixel locations in each image ($j = 1,2,…,N$). Here $A_{ij}$ is the background intensity, $B_{ij}$ is the modulation amplitude, $\theta_j$ is the angular phase information and $\delta_i$ is the phase-shift amount of each frame.

We assume that the background intensity and modulation amplitude doesn't have intraframe variation, which is a reasonable assumption under stable imaging conditions. Then, Eq (1) can be expressed as

$$I_{ij}^t = a_j + b_j \cos(\delta_j) + c_j \sin(\delta_j)$$

by defining $a_j = A_{ij}$, $b_j = B_{ij}\cos(\Phi_j)$ and $c_j = -B_{ij}\sin(\Phi_j)$. Using $M$ images and $N$ pixels we can use an over-determined least-squares method to solve for the unknown variables. The least-squares error $S_j$ can be written as

$$S_j = \sum_{\{i=1\}}^{M} (I_{\{ij\}}^t - I_{\{ij\}})^2 = \sum_{\{i=1\}}^{M} (a_j + b_j \cos(\delta_i) + c_j \sin(\delta_i) - I_{\{ij\}})^2$$

where $I_{\{ij\}}$ is the experimentally measured intensity of the interferogram.

Least-squares criteria can be written as

$$\partial S_j/\partial a_j = 0, \quad \partial S_j/\partial b_j = 0, \quad \partial S_j/\partial c_j = 0$$

which yields

$$X_j = A^{-1}/B_j$$

where

$$A = \begin{bmatrix} M & \sum_{i=1}^{M} \cos(\delta_i) & \sum_{i=1}^{M} \sin(\delta_i) \\ \sum_{i=1}^{M} \cos(\delta_i) & \sum_{i=1}^{M} \cos^2(\delta_i) & \sum_{i=1}^{M} \cos(\delta_i)\sin(\delta_i) \\ \sum_{i=1}^{M} \sin(\delta_i) & \sum_{i=1}^{M} \cos(\delta_i)\sin(\delta_i) & \sum_{i=1}^{M} \sin^2(\delta_i) \end{bmatrix}$$

$$\{X_j\} = \{a_j \quad b_j \quad c_j\}^T$$

$$\{B_j\} = \left\{\sum_{i=1}^{M} I_{ij} \quad \sum_{i=1}^{M} I_{ij}\cos(\delta_i) \quad \sum_{i=1}^{M} I_{ij}\right\}$$

From these equations, the unknowns $a_j$, $b_j$I and $c_j$ can be solved using

$$\phi_j = \tan^{-1}\left(-\frac{c_j}{b_j}\right)$$

Also, the amplitude term is found as,

$$D_j = \sqrt{\frac{\left[\frac{\left(\sum_{i=1}^{N} I_{ij}\right)}{N} - \sqrt{\Delta}\right]}{2}}$$

where

$$\Delta = \frac{\sum_{i=1}^{N} I_{ij}}{N}$$

# Analytical calculation of microbead phase distribution

Phase distribution of the microbead is calculated as follows:

$$\phi_I(x, y) = k_n \Delta_I(x, y) + k[r - \Delta_{II}(x, y)]$$

Where $n$ denotes the refractive refractive index, $k_n \Delta(x, y)$ denotes the phase delay introduced by the bead, and $k[r - \Delta_{II}(x, y)]$ denotes the phase delay introduced by the free space below microbead.

$$\Delta_I(x, y) = r - r\left[1 - \sqrt{1 - \frac{x^2 + y^2}{r^2}}\right] = \sqrt{r^2 - (x^2 + y^2)}$$

Similarly,

$$\Delta_I(x, y) = r + (-r)\left[1 - \sqrt{1 - \frac{x^2 + y^2}{r^2}}\right] = \sqrt{r^2 - (x^2 + y^2)}$$

Total thickness is determined as:

$$\Delta(x, y) = 2\sqrt{r^2 - (x^2 + y^2)}$$

Total phase delay is calculated as:

$$\phi(x, y) = k_n \Delta(x, y) + k[r - \Delta(x, y)]$$

# Noise filtering methods

In order to reduce the noise content of the reconstructed depth maps, different fltering techniques are applied and the obtained results are compared. Particularly, we compared the wavelet-based filtering approaches and a BM3D based filtering approach. Our approach was to apply these filters onto individual interferograms and obtain a reconstruction from the filtered set of interferograms. To demonstrate the validity of our noise reduction method we estimate the depth maps for a reference slide (Malvern, PVS 5113) and reference beads (Thermo Scientific, 4K100). These reference objects have pre-defined shapes and are suitable for measuring the error rates of the obtained depth maps. While measuring the noise content of the cell cultures we had to follow a different technique since the ground truth depth information was not available. To measure the effectiveness of the noise reduction method, for each observation we capture the culture twice; once with the medium flow and once after the medium flow is stopped and the remaining medium is withdrawn from the system. This allowed us to evaluate the noise introduced by the fluid flow. After the observation the medium flow is started again so as not to disturb the dynamic culture.

# Statistical analysis

To quantify agreement between observed and calculated microbead phase distribution as a measure of the error margin of the proposed methodology, the coefficient of determination was calculated as follows:

$$R^2 = 1 - \frac{\Sigma(y_{i,observed} - y_{i,calculated})^2}{\Sigma(y_{i,observed} - \bar{y})^2},$$

where

$$\bar{y} = \frac{1}{N}\sum y_{i,observed}$$

The sum operation was carried over the two-dimensional data obtained from the phase distribution calculation and the from cellular vibration observations. Statistical analyses were performed on the MATLAB software. We performed one-way ANOVA for comparisons between multiple groups with the null hypothesis that the mean values are the same.

# Acknowledgements

This work was supported by TUBITAK with project no. 116E743.